\documentclass[preprint,authoryear,12pt]{elsarticle}

\usepackage[T1]{fontenc}
\usepackage[english]{babel}
\usepackage{graphicx}
\usepackage{subfig}
\usepackage{amsmath}
\usepackage{amssymb}
\usepackage{bm}
\usepackage{url}
\usepackage{multicol}
\usepackage{tabularx,ragged2e}
\usepackage[ruled,vlined]{algorithm2e} 
\usepackage{bbm}
\usepackage{adjustbox}
\usepackage{lineno} 
\usepackage{hyperref}
\usepackage[nameinlink]{cleveref}

\hypersetup{
    colorlinks  = true,
    linkcolor   = black,
    urlcolor    = black,
    citecolor   = blue
}

\newcommand{\mytilde}[1]{
    \mathrel{\overset{\makebox[0pt]{\mbox{\normalfont\tiny #1}}}{\sim}}
}

\newenvironment{nalign}{
    \begin{equation}
    \begin{aligned}
}{
    \end{aligned}
    \end{equation}
    \ignorespacesafterend
}

\usepackage[normalem]{ulem}
\usepackage{xcolor}

\journal{Computational Statistics and Data Analysis}

\begin{document}

\begin{frontmatter}
    
\title{Gaussian graphical modeling for spectrometric data analysis}

\author[polimi]{Laura Codazzi}
\author[polimi]{Alessandro Colombi}
\author[polimi]{Matteo Gianella}
\author[unicatt]{Raffaele Argiento}
\author[unicatt]{Lucia Paci}
\author[unicatt]{Alessia Pini}

\affiliation[polimi]{%
    organization={Department of Mathematics, Politecnico di Milano},
    addressline={Piazza Leonardo da Vinci 32}, 
    city={Milan},
    postcode={20133}, 
    country={ITALY}
}

\affiliation[unicatt]{%
    organization={Department of Statistical Sciences, Università Cattolica del Sacro Cuore},
    addressline={Largo Agostino Gemelli 1}, 
    city={Milan},
    postcode={20123}, 
    country={ITALY}
}

\begin{abstract}
    Motivated by the analysis of spectrometric data, we introduce a Gaussian graphical model for learning the dependence structure among frequency bands of the infrared absorbance spectrum.
     The spectra are  
    modeled as continuous functional data through a B-spline basis expansion and
    a Gaussian graphical model is assumed as a prior specification for the smoothing coefficients to induce sparsity in their precision matrix. 
    Bayesian inference is carried out to simultaneously smooth the curves and to estimate the  conditional independence structure between portions of the functional domain. 
    The proposed model is applied to the analysis of  infrared absorbance spectra of strawberry purees.
\end{abstract}




\end{frontmatter}


\section{Introduction}
\label{intro}

The analysis of the interaction of infrared radiation with matter by absorption, emission, or reflection is accomplished by \textit{infrared spectroscopy}. This technique is used to study and  identify chemical substances in solid, liquid, or gaseous forms. 
For instance, mid-infrared spectroscopy coupled with chemometrics or statistical techniques have  been used in the literature to study the substance composition and detect the presence of adulterants in food  
\citep{Kemsley1996,downey1997,holland.kemsley.wilson1998,MezaMrquez2010}. 

From a mathematical point of view, a spectrum is a continuous function of the wavelength. 
The dependence structure between the signal at different wavelength bands is important to understand which bands are related to the different components: 
if two different bands of the spectrum are associated, we can conclude that they refer to the same components, or to closely related components. Our goal in this paper is then to investigate the structure of conditional dependence among different portions of an absorbance spectrum.

Since infrared spectra are continuous functions of the wavelength, we embed this problem in the framework of functional data analysis \citep[see e.g.,][]{lee2002estimating,ramsay.silverman2005,xiao2016fast}. In the Bayesian setting, \citet{yang2016} proposed a  hierarchical model with Gaussian-Wishart processes for simultaneously smoothing multiple functional observations and estimating mean-covariance functions. However, like any Gaussian process  based approach, their model suffers serious computational burden when functional data are observed on high-dimensional grids. To address this computational issue, \citet{yang2017} proposed to approximate the underlying true functional data  with basis functions, and derive the induced Bayesian hierarchical model for the associated smoothing coefficients based on  a Gaussian-Wishart prior. 

Following the latter approach, we interpret the analysis of spectrometric data as a smoothing problem of functional data, followed by inference on the smoothing coefficients. Specifically, we use a B-spline basis expansion to represent the functional data: indeed, a generic B-spline basis function has a compact support, meaning that it is different from zero only in a portion of the domain, i.e., on a defined portion of the spectrum. If the number of basis functions is large, this portion is very small compared to the whole domain, and identifies a very precise band of the spectrum. 
Through the basis expansion, a coefficient is associated to every basis function, and the whole function can be represented through the vector of associated coefficients $\bm{\beta}$. As a consequence, we can detect a hidden association between different bands of the spectrum through an association between the corresponding smoothing coefficients.

Inference on B-spline basis coefficients has been used in the functional data analysis literature for assessing the mean structure of functional data along the domain \citep[see e.g.,][]{pini2016interval}. Here, instead, we focus on the dependence structure of the smoothing coefficients. With a Bayesian perspective, this concerns the specification of a prior distribution for the coefficients that encodes a dependence structure. Usually, a random walk prior  is assumed to induce smoothness on the $\bm{\beta}$'s  \citep{lang2004,crainiceanu2010,telesca2008}. First and second order random priors are typically used, that translate into a banded precision penalization matrix that is fixed and depends only on a smoothing parameter. However, in our application, assuming that the dependence structure is known and restricted
to the structure induced by a random walk prior appears inadequate. Indeed, coefficients associated with closer bands of the spectrum are not necessarily expected to be similar since they may relate to different components. Rather, long range interactions may also exist between far apart portions of the spectrum. Hence, a more flexible modeling framework 
is advocated.

In this paper, we move a step forward and develop a method to learn the dependence structure among portions of the absorbance spectrum based on graphical modeling. 
Graphical models describe a mapping between a graph and a family of multivariate probability models \citep{lauritzen1996}. Namely, they embody conditional independence relationships among a set of random variables which can be read off from the graph. Usually, the structure of the underlying graph is unknown and needs to be estimated on the basis of the available data: this is referred to as graph structural learning. 

Graphical models have been  widely
applied  to infer several types of
networks, particularly under the Bayesian framework, see among others \citet{dobra.hans.jones.nevins.yao.west2004,peterson.stingo.vannucci2015,tan.jasra.deiorio.ebbels2017,Ni2017,cremaschi2019,PACI2020106880}. 
All the cited works assume a graphical model as a model for the data, i.e., the graph represents the conditional independence relationships among the observed variables. Here, instead, we introduce a graphical model  as a prior specification for the smoothing coefficients, such that the graph interprets the dependence structure of the portions of the absorbance spectrum. 
Graph-based priors for regression coefficients have been studied in variable selection setting by, among others, \citet{liu2014}, \citet{chakraborty2019} and \citet{cai2020}, who proposed Bayesian regularization priors to favor sparsity and clustering based on the graph Laplacian. However, these approaches are not suitable to deal with functional data where curve-specific coefficients are needed to simultaneously smooth the curves.

In addition, recent works focused on graphical modeling for multivariate functional data analysis  \citep{zhu2016,li2018,qiao2019,qiao2020}. These approaches, known  as functional graphical models, aim at depicting the conditional dependence structure among multiple random functions observed over a set of individuals, i.e., they assume a graphical model  where the vertices of the underlying graph are the multiple functions and the edges are estimated from the conditional covariance function extended to the functional domain. Rather, in our setting, we work with univariate functional data modeled through a  B-spline basis expansion and we assume a graphical model where the nodes of the graph are the smoothing coefficients and the edges reflect  the conditional dependence structure among the portions of the functional domain associated to such coefficients.

Summarizing, our contribution is to introduce a Bayesian hierarchical model for simultaneously smoothing functional data and learning the independence structure along the domain of the curves. Absorbance spectra are modeled as continuous functional data through a cubic B-spline basis expansion  such that the  dependence between two bands of the spectrum is reflected
in the relationship among the corresponding smoothing coefficients. A Gaussian graphical model is then assumed as a prior for basis expansion coefficients to enable structural learning of frequency bands of the spectrum. 
To the best of our knowledge, this is the first work where a Gaussian graphical model is assumed as a prior model for the smoothing  coefficients in the analysis of functional data. 
On the computational side, we design an efficient sampling strategy to approximate the joint posterior distribution of the graph and model parameters. 
We illustrate our method on simulated datasets as well as on a real data set, studying the infrared absorbance spectra of strawberry purees. 

The remainder of this paper is organized as follows. \Cref{sec:FGM} introduces the Bayesian graphical model for functional data analysis, discusses the smoothing procedure as well as the covariance selection model. 
In  \Cref{sec:sampling} we describe the sampling strategy implemented to sample from the joint posterior  distribution of all model parameters, including the graph, and discuss how to summarize the output.  \Cref{sec:simulation} presents a simulation study, while  \Cref{sec:analysis} illustrates the results of a real data analysis carried out to study the infrared absorbance spectra of strawberry purees. Finally we conclude with a brief discussion in  \Cref{sec:conclusion}.

\section{Bayesian graphical model for smoothing functional data} 
\label{sec:FGM}
\subsection{Smoothing procedure}
As we have anticipated in the Introduction, we reinterpret our task as a traditional smoothing procedure of functional data 
in a Bayesian framework. Let $n$ be the number of curves, $r$ the number of grid points, i.e., the wavelengths at which absorbance is measured, and $p$ be the number of basis functions chosen for the smoothing.
Let $y_{i}(s)$ be the absorbance at wavelength $s\in[l,u]$ for unit $i$, $i=1,\dots,n$, where $[l,u]\subset\mathbb{R}$ is the common domain of all curves. The whole curve $y_{i}(s)$ is then the absorbance spectrum of unit $i$. According to the usual smoothing technique, the model we assume for the $i$-th spectrum is:
\begin{equation*}
    y_{i}(s) = \sum_{j=1}^{p}\beta_{ij}\varphi_{j}(s) +\varepsilon_{i}(s),
    \label{smoothing_equation}
\end{equation*}
where $\varphi_{1},\dots,\varphi_{p}$ are suitable cubic B-spline basis functions, $\bm{\beta}_{i}= (\beta_{i1},\dots,\beta_{ip})^\top$ is the spline coefficients’ vector specific to $i$-th  curve and $\varepsilon_{i}(s)\mytilde{iid}N(0,\tau^{2}_{\varepsilon})$.
We introduce the cubic B-spline design matrix $\bm{\Phi}\in\mathbb{R}^{r \times p}$ as:
\begin{equation}
    \bm{\Phi} =
	\begin{bmatrix}
	    \varphi_{1}\left(s_{1}\right) &\varphi_{2}\left(s_{1}\right) &\dots &\varphi_{p}\left(s_{1}\right)\\
	    \varphi_{1}\left(s_{2}\right) &\varphi_{2}\left(s_{2}\right) &\dots &\varphi_{p}\left(s_{2}\right)\\
    	\vdots &\vdots &\ddots &\vdots\\
    	\varphi_{1}\left(s_{r}\right) &\varphi_{2}\left(s_{r}\right) &\dots &\varphi_{p}\left(s_{r}\right)\\
	\end{bmatrix}.
	\label{eq:design_matrix}
\end{equation}
So, the element $lj$-th of $\bm{\Phi}$ is $j$-th basis function $\varphi_j(\cdot)$ evaluated at the $l$-th grid point $s_l$. 
Note that, the smoothing model can be also used in the case where curves are measured at different wavelengths, by modifying $\bm{\Phi}$ accordingly. 

\begin{figure}[ht]
    \centering
    \includegraphics{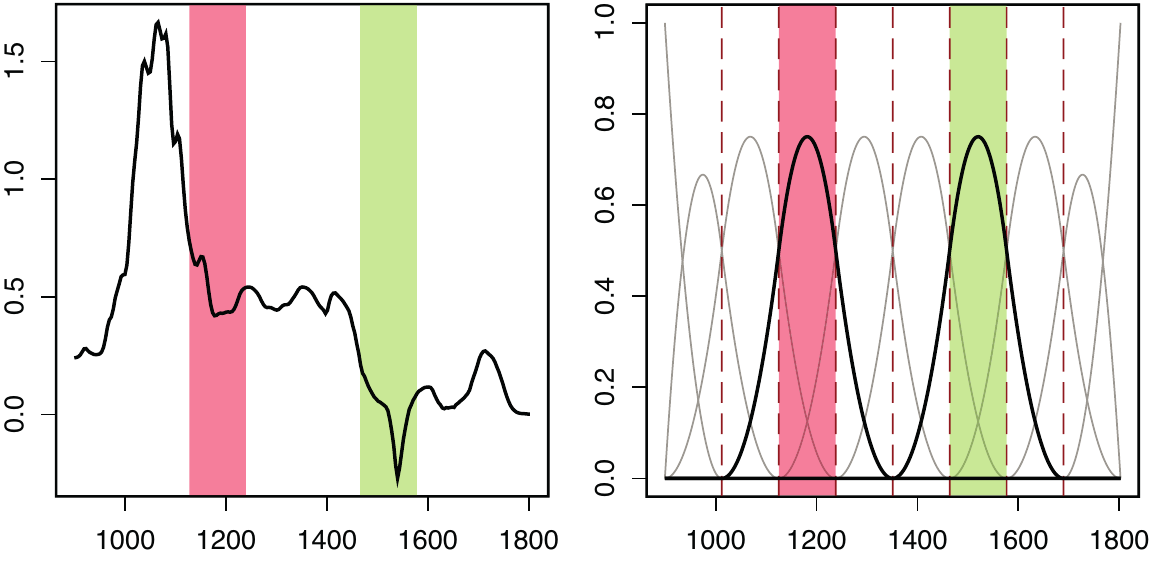}
    \caption{Example of a curve (left panel) and the corresponding  10-dimensional cubic B-spline basis (right panel).}
    \label{fig:splines}
\end{figure}

The right panel of \Cref{fig:splines}  shows an example of a 10-dimensional cubic B-spline basis for the curve in the left panel. Note that 
the support of each basis function is a small portion of the domain, which decreases its size when the number of basis functions increases. For instance, in \Cref{fig:splines}, the support of each basis function overlaps with only four other function' supports. To facilitate the interpretation of the results of our model, we will consider each B-spline basis coefficient as representative of the central part of each B-spline basis function only. 

Let $\bm{Y}_i = \left(y_i(s_1), \dots, y_i(s_r) \right)^\top$ be the absorbance spectrum at all observed wavelengths of curve $i$. The smoothing model assumes the observation to follow a conditional normal sampling distribution, that is
\begin{equation}
    \bm{Y}_{i}~\lvert~\bm{\beta}_{i},~\tau^{2}_{\varepsilon} ~\mytilde{ind}~N_{r}\left(\bm{\Phi}\bm{\beta}_{i},~\tau^{2}_{\varepsilon}\mathbf{I}_{r}\right),
    \label{y_conditional_equation}
\end{equation}
where $\tau^2_\epsilon$ is the error variance and $\bm{I}_r$ is the $r\times r$ identity matrix.

In this work, we rely on a hierarchical Bayesian approach to smooth all
functional observations simultaneously,  enabling the borrowing of strength across all curves. To accomplish that, we place a prior distribution for the $\bm{\beta}_{i}$ coefficients given by 
\begin{equation}
    \bm{\beta}_{1},\dots,\bm{\beta}_{n}~\lvert~\bm{\mu},\bm{\Omega}\mytilde{iid}N_{p}(\bm{\mu},\bm{\Omega}^{-1}),
    \label{beta_conditional_equation}
\end{equation}
where $\bm{\mu}$ is the prior mean vector and $\bm{\Omega}$ is the precision matrix.  \citet{yang2017} derived the conjugate prior distribution for $\bm{\mu}$ and $\bm{\Omega}$ from the assumption that the underlying true functional data come from a Gaussian process, and thus resulting in a Normal-Wishart distribution that depends on the basis functions. In this work, instead, we frame the prior model for the $\bm{\beta}$'s into the Gaussian graphical modeling setting. Indeed, given the peculiar form of each B-spline function, the dependence between two bands of the spectrum is reflected in the relationship among the corresponding smoothing regression parameters. Hence, our goal boils down to the study of the conditional dependence structure of the $\bm{\beta}$’s encoded in the precision matrix $\bm{\Omega}$.

\subsection{Graph structural learning}

\begin{figure}[!ht]
    \centering
    \includegraphics{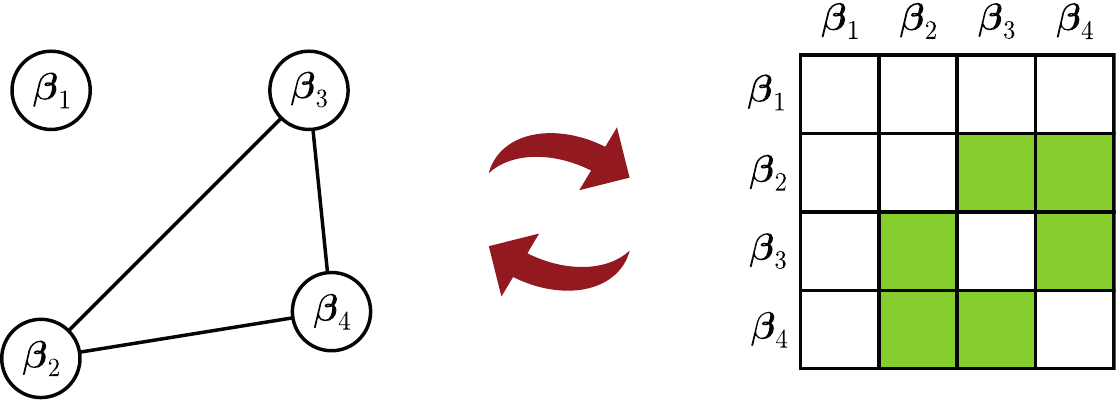}
    \caption{Example of a Gaussian graphical model with four nodes. On the left, the graph with four nodes and, on the right, the corresponding precision matrix where green cells represent nonzero entries.}
    \label{fig:graph_matrix_relation}
\end{figure}

%
%
Let $G = (V,E)$ be an undirected graph defined by the node set $V = \{1,\dots,p\}$ representing the coefficients and by the undirected edge set $E \subset V \times V$. We assume the precision matrix $\bm{\Omega}$ to be  Markov with respect to $G$. Namely, dropping the subscript $i$ for simplicity in the exposition, coefficients $\beta_{j}$ and $\beta_{k}$ are conditionally independent given all the remaining variables,  $\bm{\beta}_{\backslash\{j,k\}}$, whenever $\{j,k\}\not\in E$, if and only if the corresponding entry in matrix $\bm{\Omega}$ is zero, i.e., $\beta_{j} \perp \beta_{k}~\lvert~\bm{\beta}_{\backslash\{j,k\}} \Leftrightarrow \bm{\Omega}_{jk}=0$; see \Cref{fig:graph_matrix_relation} for an example.
The graph $G$ summarizes the conditional independence structure of the coefficients and represents the main goal of our inference. Indeed, the graph is unknown and must be estimated. 
Note that this is different from estimating the conditional network among the predictors of a regression model as in \citet{peterson2016}. 
Unlike measurement-error models in which predictors and response are both measured
with error, spline basis expansion are fixed and known.  Hence, we infer a sparse network on the associated smoothing coefficients, which provides information on which basis have similar slope.

Given $\mathbb{P}_G$ the space of $p\times p$ positive definite matrices that are Markov with respect to graph $G$, we assume the precision matrix $\bm{\Omega} \in \mathbb{P}_G$ to be distributed, a priori, as a $\operatorname{G-Wishart}(d,\bm{D})$ \citep{roveratoa.2002,letac2007} with density 
\begin{equation}
    P\left(\bm{\Omega}~\lvert~ G, d,\bm{D} \right) = I_{G}\left(d, \bm{D}\right)^{-1} \lvert \bm{\Omega} \rvert^{(d - 2)/2} \exp\left\{- \frac{1}{2}\mathrm{tr}\left(\bm{\Omega}\bm{D}\right)\right\},
    \label{eqn:Gwish}
\end{equation}
where $d > 2$ is the shape parameter, 
$\bm{D}$ is the $p\times p$ positive definite inverse scale matrix,  $\text{tr}(\cdot)$ denotes the trace operator and  $I_{G}(d, \bm{D})$ is the normalizing constant,
\begin{equation}
I_G(d,\bm{D}) = \int_{\mathbb{P}_G}\left|\bm{\Omega}\right|^{(d - 2)/2}\exp\left\{- \frac{1}{2}\mathrm{tr}\left(\bm{\Omega}\bm{D}\right)\right\}d\bm{\Omega}.
\label{eq:normalizing}
\end{equation}
The main challenge associated with density in  \eqref{eqn:Gwish} arises from the normalizing constant in equation \eqref{eq:normalizing}; although an analytic form does exist 
\citep{uhler2018}, the expression is mathematically complex and very difficult to compute in practice. 
The sampling strategy described in \Cref{sec:sampling} is then  designed to avoid the calculation of such intractable normalizing constant.


The last ingredient to fully specify the model is the prior distribution on the graph $G$. Several alternatives have been proposed in the literature, see among others \cite{dobra.hans.jones.nevins.yao.west2004, jonesb.carvalhoc.dobraa.hansc.carterc.westm.2005, scottj.gcarvalhoc.m2008,PACI2020106880}. 
In this work, we 
consider a discrete uniform distribution over the space of all possible graphs $\mathcal{G}$, that is
\begin{equation}
    \pi(G) = \frac{1}{|\mathcal{G}|},\text{ for each } G \in \mathcal{G}.
    \label{eqn:prior_unif}
\end{equation}
Alternatively, it is possible to induce a prior on the graph by assigning to each possible edge $f\in E$ an independent Bernoulli prior with parameter $\theta_{f}\in(0,1)$, that yields
\begin{nalign}
	\pi(G) ~&\propto~\prod_{f \in E\vphantom{\bar{E}}}\theta_{f}\prod_{f \in \bar{E}}(1-\theta_{f}),
	\label{eqn:Bernoullipr}
\end{nalign}
where $\bar{E}$ is the complement of $E$. 
In a general setting, the Bernoulli parameters $\theta_f$ can be different from edge to edge. Conversely, when $\theta_{f} = \theta,~\forall{f \in E}$, we get $\pi(G) \propto \theta^{|E|}(1 - \theta)^{\binom{p}{2}-|E|}$, where $|E|$ is the cardinality of $E$. 
Notice that, when $\theta_{f} = 0.5, ~\forall{f \in E}$, prior \eqref{eqn:Bernoullipr} collapses to the uniform prior in equation  \eqref{eqn:prior_unif}.

For the remaining parameters, we choose independent  semi-conjugate vague priors. Summing up, our Bayesian hierarchical model is defined as follows:
\begin{nalign}
    \bm{Y}_{i}~\lvert~\bm{\beta}_{i},~\tau^{2}_{\varepsilon} &~\mytilde{ind}~\mbox{N}_{r}\left(\bm{\Phi}\bm{\beta}_{i},\tau^{2}_{\varepsilon}\mathbf{I}_{r}\right) \\
	\bm{\beta}_{1},\dots,\bm{\beta}_{n}~\lvert~\bm{\mu},~\bm{\Omega}&~\mytilde{iid}~\mbox{N}_{p}\left(\bm{\mu},\bm{\Omega^{-1}}\right)\\
			\bm{\Omega}~\lvert~ G &~\mytilde{~}~\operatorname{G-Wishart}\left(d,\bm{D}\right) \\
	G&~\mytilde{~}~\pi\left(G\right)\\
	\bm{\mu}&~\mytilde{~}~\mbox{N}_{p}\left(\bm{0},\sigma_{\mu}^{2}\mathbf{I}_{p}\right)\\
		\tau_{\varepsilon}^2 &~\mytilde{~}~ \operatorname{IG}\left(a,b\right),
	\label{fgm_model}
\end{nalign}
\noindent where $\operatorname{IG}(a,b)$ denotes the Inverse Gamma distribution with shape parameter $a$ and rate parameter $b$. 
The Bayesian model in equation  \eqref{fgm_model} enables joint modeling of all the curves with borrowing strength of information across functional data as well as to fully account for uncertainty over the graph.

\section{Sampling strategy} 
\label{sec:sampling}
To deliver Bayesian posterior inference we need to approximate the joint posterior distribution of the $\bm{\beta}$'s, $\tau^2_\varepsilon$, $\bm{\mu}$, $\bm{\Omega}$ and $G$. 
Here, one of the main challenge arises from the graph $G$ that belongs to a huge discrete space endowed with a complex topology. 
Indeed, there are $2^{p(p-1)/2}$ possible graphs representing the conditional independence structure of a $p$-dimensional vector of coefficients.
For instance, with $p=7$ B-spline basis functions only, we end up with more than $2$ million graphs. As consequence,  any simulation algorithm  struggles to explore the space of all possible graphs.


%
A customary approach to explore such space is by means of a  Monte Carlo Markov chain (MCMC) algorithm. In this framework,  a reversible Markov chain is built such that its limiting distribution is the joint posterior of the graph and the precision matrix. Usually, the algorithm proceeds by proposing at each step a new graph which differs from the current one for only one edge, i.e., local moves. This procedure is called add-delete Metropolis Hastings, see among others \citet{giudicip.castelor.2003, bhadraa.mallickb.k.2013}. Then, the Markov chain moves over the graph space by comparing different graphs, in order to find the one having the highest marginal posterior.
However, this procedure has some drawbacks. For instance, 
\citet{jonesb.carvalhoc.dobraa.hansc.carterc.westm.2005} showed that the add-delete Metropolis Hastings methods suffer from lack of convergence  in high-dimensional settings. 

As an alternative, \citet{mohammadi.wit2015} proposed a birth and death MCMC algorithm (BDMCMC) based  on a continuous time Markov
process where  jumps between birth and death events are taken to be random variables with  specific rates. The authors showed that the BDMCMC outperforms alternative Bayesian approaches in terms of convergence and mixing in the graph space as well as computing time.

We build a sampling strategy that blends Gibbs-sampler steps for updating the smoothing components of the Bayesian model with a BDMCMC step for updating the precision matrix $\bm{\Omega}$ and the graph $G$, as describe in Algorithm \ref{algo}. As concerns the Gibbs-sampler steps,  
we sample the $\bm{\beta}$ coefficients from their full conditional distribution given by
\begin{equation}
    \bm{\beta}_{i}~\lvert~\mathrm{rest}  \sim N_{p}\left(\bm{b}_{n_{i}},		\bm{B}_{n}\right),
\label{eqn:full_beta}
\end{equation}
where $\bm{B}_{n} = \left(\bm{\Phi}^{T}\bm\Phi/\tau_{\varepsilon}^{2} + \bm{\Omega}\right)^{-1}$ and $\bm{b}_{n_{i}} = \bm{B}_{n}\left(\bm\Phi^{T}\bm{Y}_{i}/\tau_{\varepsilon}^{2} + \bm{\Omega}\bm{\mu}\right)$ for all  $i=1\dots n$. The full conditional distribution of $\bm{\mu}$ is given by
\begin{equation}
\bm{\mu}~\lvert~\mathrm{rest} \sim N_{p} \left(\bm{m}, \bm{M}\right),
\label{full_mu}
\end{equation}
where   $\bm{M}=\bm{I}_{p}/\sigma_{\mu}^{2} + n\bm{\Omega}$ and $\bm{m} = \bm{M}^{-1} \bm{\Omega} \sum_{i = 1}^{n} \bm{\beta}_{i}$. For the $\tau_{\varepsilon}^{2}$, we sample from 
\begin{equation}
\label{full_tau}
\tau_{\varepsilon}^{2}~\lvert~ \mathrm{rest} \sim \mathrm{IG}\left(\dfrac{nr + a}{2},\frac{1}{2}\left( b + \sum_{i = 1}^{n} \left(\bm{Y}_{i} - \bm\Phi\bm{\beta}_{i}\right)^{T}\left(\bm{Y}_{i} - \bm\Phi\bm{\beta}_{i}\right)\right)\right).
\end{equation}
Finally, since only $\bm{\beta}$'s are directly influenced by the graphical component, we can rewrite 
the joint full conditional of  $(G,\bm{\Omega})$ as 
\begin{equation}
 P(G,\bm{\Omega}\lvert\bm{Y}_{1},...,\bm{Y}_{n},\tau_{\varepsilon}^{2},\bm{\mu},\bm{\beta}_{1},...,\bm{\beta}_{n}) = P\left(G,\bm{\Omega}\left|\right.\bm{\beta}_{1},...,\bm{\beta}_{n},\bm{\mu}\right).
 \label{eq:omegaG_fc}
\end{equation}
As introduced in \Cref{sec:FGM}, 
even when the graph is known, sampling from a G-Wishart distribution poses computational issues. 
As a consequence, any method based on the computation of the marginal full conditional of the graph,
\begin{equation}
    P(G ~\lvert~ \bm{\beta}_{1},...,\bm{\beta}_{n} ) \propto \pi(G)\frac{I_{G}\left(d+n, D+U\right)}{I_{G}\left(d, D\right)},
    \label{eq:marginalG}
\end{equation}
where $U = \sum_{i = 1}^{n}(\bm{\beta}_{i}-\bm{\mu})(\bm{\beta}_{i}-\bm{\mu})^{T}$,  is  not efficient because requires the computation of both the prior and posterior normalizing constant of density \eqref{eqn:Gwish}. 

To   sample from \eqref{eq:omegaG_fc} while avoiding  computation of \eqref{eq:marginalG}, we employ the BDMCMC approach developed by \citet{mohammadi.wit2015} for Gaussian graphical model selection. The BDMCMC algorithm is based on a continuous time Markov process,  where 
edges are added or removed via birth or death events and the  time between jumps is taken to be an exponential random variable with specific rates. Birth and death events are characterized as follows:
\begin{itemize}
    \item Death: Each edge $e = (j,k)$ between nodes $j$ and $k$, where $j<k$, dies independently of the others as a Poisson process with rate $\delta_{e}(\bm{\Omega})$. The overall death rate is then $\delta(\bm{\Omega})=\sum_{e\in E}\delta_{e}(\bm{\Omega})$.
When the death of an edge $e$ occurs, then the process jumps to a new state $(G^{-e},\bm{\Omega}^{-e})$, where $G^{-e}=(V,E\backslash\{e\})$ and $\bm{\Omega}^{-e}\in P_{G^{-e}}$. 
\item Birth: A new edge $e = (j,k)$ between nodes $j$ and $k$, where $j<k$, is born independently of the others as a Poisson process with birth rate $\lambda_{e}(\bm{\Omega})$. The overall birth rate is then $\lambda(\bm{\Omega}) = \sum_{e\in \bar{E}}\lambda_{e}(\bm{\Omega})$. If the birth of an edge $e\in \bar{E}$ occurs, then the process jumps to a new state $(G^{+e}, \bm{\Omega}^{+e})$, where $G^{+e}=(V,E \cup \{e\})$ and $\bm{\Omega}^{+e}\in P_{G^{+e}}$.

\end{itemize}
Birth and death processes are assumed to be independent. Hence, the intensity of the jump process is given by $w(\bm{\Omega}) = 1/\left(\lambda(\bm{\Omega}) + \delta(\bm{\Omega})\right)$. We recall that this implies that the time between two consecutive jumps is exponentially distributed with mean $w(\bm{\Omega})$ that we refer to as the weight of the state. 
The choice of birth and death rates is made such that the balance conditions given by \citet{c.j.preston} hold, so that the stationary distribution of the birth-death  process  is  the  full conditional  distribution in \eqref{eq:omegaG_fc}. 
The last step is then updating the precision matrix $\bm{\Omega}$ using the exact sampler described in \citet{mohammadi.wit2015} and originally developed by \citet{alexlenkoski2013}. 
The BDMCMC for customary Gaussian graphical model selection is implemented by the \texttt{R} package \texttt{BDgraph} \citep{BDgraph}.

\begin{algorithm}
\SetAlgoLined
For each iteration:\
\begin{itemize}
    \item[\textbf{Step 1.}] Sample the regression parameters
    \begin{itemize}
        \item [\textbf{1.1}] Sample the parameters $\bm{\beta}$'s from their full conditional in \eqref{eqn:full_beta}.
      \item [\textbf{1.2}] Sample the parameter $\mu$ from its full conditional in \eqref{full_mu}.
      \item [\textbf{1.3}] Sample the parameter $\tau_{\varepsilon}^{2}$ from its full conditional in \eqref{full_tau}.
    \end{itemize}
    \item [\textbf{Step 2.}] Sample $(G,\bm{\Sigma}^{-1})$ from their full conditional using the BDMCMC algorithm and save the expected holding time $w(\bm{\Omega})$. \end{itemize}
    \caption{MCMC sampler.}
    \label{algo}
\end{algorithm}

To sum up, our sampling strategy is given in Algorithm \ref{algo}. As far as Step 1 is concerned, it consists in sampling from conjugate full conditionals described above. Then we exploit the \texttt{BDgraph} function to complete Step 2. We call the function for only one iteration using as argument the parameters updated at Step 1. In other words, we run the \texttt{BDgraph} function as if data were $\bm{\beta_{1}}-\bm{\mu},...,\bm{\beta_{n}}-\bm{\mu}$. We also create an auxiliary structure to store all the explored graphs  and its associated expected holding time $w(\bm{\Omega})$.  The \texttt{R} code implementing Algorithm \ref{algo} is available under request.

\subsection{Posterior graph}
\label{sec:graph_est}
Given a MCMC output of size $T$,  posterior summaries of  
the precision matrix and 
the graph need to take into account the fact that Step 2 of Algorithm \ref{algo} involves sampling from a continuous time Markov process. In particular, the posterior estimate of $\bm{\Omega}$ is based on the Rao-Blackwell theorem \citep{cappeo.roberc.prydent.2003}, yielding to a posterior estimate that is
\begin{equation}
    \hat{\bm{\Omega}} = \frac{\sum_{t=1}^T w_{t}(\bm{\Omega})\bm{\Omega}_{t}}{\sum_{t=1}^T w_{t}(\bm{\Omega})}.
\end{equation}

As far as the posterior graph is concerned, one approach for selecting the graph is to use the maximum a posteriori estimate, i.e., the highest posterior probability graph. However, this approach is not
generally reliable since the space of possible graphs is quite
large and any particular graph may be encountered only a few
times in the course of the MCMC sampling. A more practical
solution is instead to estimate the marginal posterior edge inclusion probabilities. Namely, given the MCMC output, the posterior inclusion probabilities are estimated as
\begin{equation}
    \hat{p}_{jk} = \frac{\sum_{t=1}^{T}\mathbbm{1}\left((j,k) \in E_{t}\right)w(\bm{\Omega}_{t})}{\sum_{t=1}^{T}w(\bm{\Omega}_{t})},
    \label{eqn:post_prob}
\end{equation}
where $\mathbbm{1}\left((j,k) \in E_t\right)$ is the indicator function representing the inclusion of the edge linking nodes $j$ and $k$ and  drawn at iteration $t$. 
Here, we select the graph containing all edges whose posterior inclusion probability in \eqref{eqn:post_prob} exceeds a given threshold $s$.  In particular, we compare two different thresholds. The first one is $s=0.5$, in analogy with the median probability model of \citet{barbieri2004}, originally proposed in the linear regression setting. The second  is based on the Bayesian False Discovery  rate (BFDR; \citealt{muller2007,peterson.stingo.vannucci2015})
\begin{equation}
    \mbox{BFDR} = \dfrac{\sum_{j < k}(1 - \hat{p}_{jk})\mathbbm{1}\left(\hat{p}_{jk} \geq s\right)}{\sum_{j < k}\mathbbm{1}\left(\hat{p}_{jk} \geq s\right)},
    \label{eqn:BFDR}
\end{equation}
where  $s$ is selected so that BFDR in equation \eqref{eqn:BFDR} is below $ 0.05$.

\section{Simulation study}
\label{sec:simulation}

We carried out a simulation study  to evaluate the ability of our methodology to recover the structure of the generating graph as well as to compare it with the state of the art. 
In particular, we analyzed three different experiments: in Section \ref{sec:sim1&2} we focus on the ability of our technique to learn the structure of the underlying graph, simulating data from two different structures. Further, Section \ref{sec:sim_comparison} shows a numerical comparison with \citet{yang2017}, in terms of both the ability of the two techniques to estimate the covariance matrix and on the smoothing performance. 

\subsection{Structural learning}
\label{sec:sim1&2}
We consider a graph with $p=40$ nodes to mimic the dimension of real data, see \Cref{sec:analysis}; then we sample a precision matrix $\bm{\Omega}$ from a $\operatorname{G-Wishart}(d=3, \bm{D}=\bm{I}_{p})$ constrained by the graph and draw the  $\bm{\beta}$ parameters from \eqref{beta_conditional_equation}. Finally, we simulate $50$ datasets from the sampling distribution in \eqref{y_conditional_equation}, where $n=200$ and $\tau^2_\epsilon = 0.01$. 

We fit model \eqref{fgm_model} on the simulated datasets and all the results are based on the output of Algorithm \ref{algo} with 60,000 iterations and a burn-in of 10,000. As far as  the choice of the hyperparameters is concerned, we set 
the prior variance of $\mu$ equal to $\sigma_\mu^2 = 100$ and the parameters of the prior distribution of variance $\tau^2_{\varepsilon}$ equal to $a = 10$ and $b = 0.001$, i.e., standard vague priors. 
We computed the edge posterior inclusion probabilities and provide a posterior estimate of the graph using the two thresholds described in Section \ref{sec:graph_est}. 

We evaluate the ability of recovering the underlying graph, using the Structural Hamming Distance (SHD, \citealt{SHD}), which represents the number of edge
insertions and deletions needed to transform the estimated graph into the true graph. Clearly, lower values of SHD
correspond to better performances. For sake of comparison the SHD has been standardized over the maximum number of edges.

\begin{figure}[ht]
    \centering
    \includegraphics{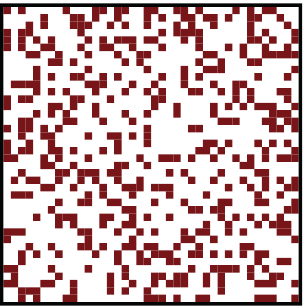}
    \caption{The true non-structured graph of simulation Experiment 1.}
    \label{fig:random_graph}
\end{figure}

\paragraph{Experiment 1: non-structured graph}
In the first experiment we consider a non-structured graph with sparsity level of $0.3$, i.e.,  each edge has a probability equal to $0.3$ of being in the graph, as shown in  \Cref{fig:random_graph}.  

\begin{figure}[ht]
    \centering
    \includegraphics{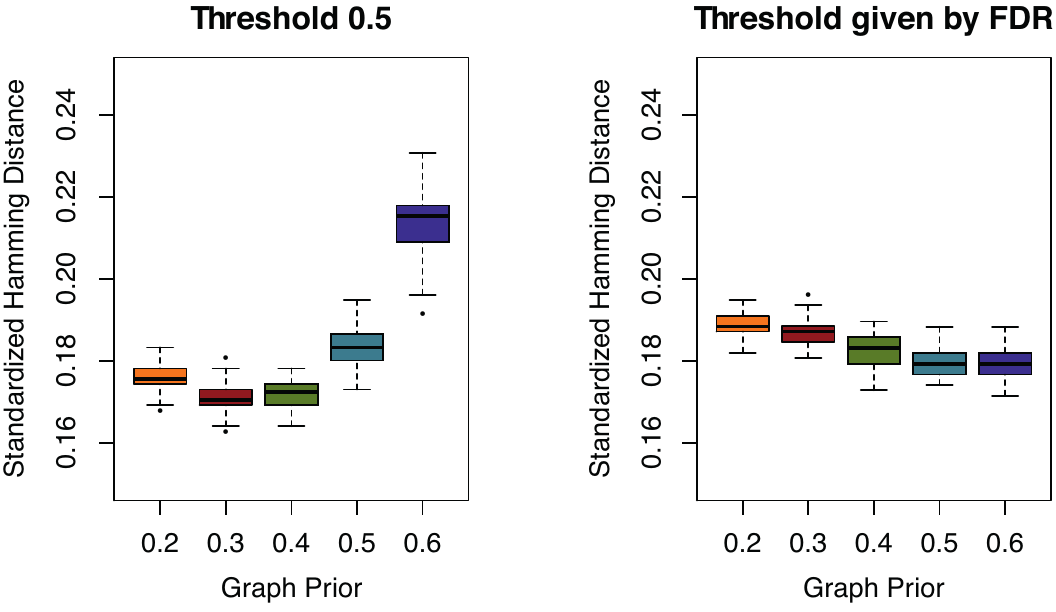}
    \caption{Boxplot of the standardized structural Hamming distances over the $50$ replicates for Experiment 1.}
    \label{fig:confrontopriorp40ber03noAUC}
\end{figure}

We fit model \eqref{fgm_model} assuming different hyperparameters for the graph prior in  \eqref{eqn:Bernoullipr}  
with prior probability $0.2$, $0.3$, $0.4$, $0.5$ and $0.6$. 
\Cref{fig:confrontopriorp40ber03noAUC} shows the boxplot of the standardized SHD values over the $50$ replicates. From the latter we note that 
the best posterior estimate of the graph is obtained when the $0.5$ threshold is employed and the graph prior probability is $0.3$, i.e., the value equal to the true sparsity level. However, the median graph is quite sensitive to prior choices while the BFDR yields to more stable results.  

\begin{figure}[ht]
    \centering
    \includegraphics{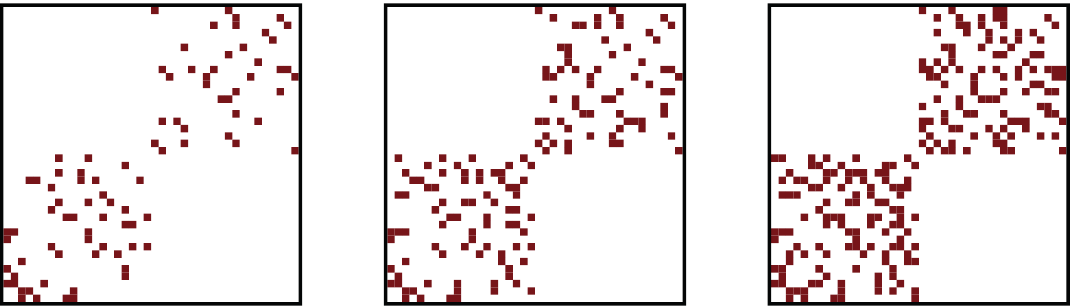}
    \caption{The true clustered graphs of Experiment 2 with sparsity levels of $0.1$, $0.2$, and $0.3$ in the left, middle and right panel, respectively.}
    \label{fig:cluster_graphs}
\end{figure}

\begin{figure}[ht]
    \centering
    \includegraphics{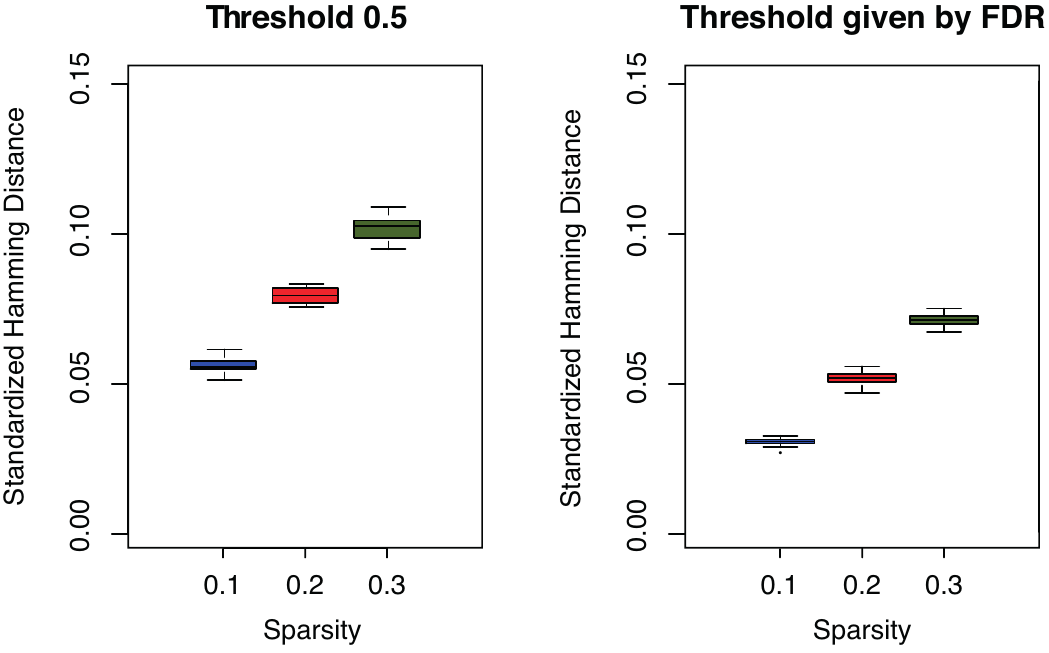}
    \caption{Boxplot of the standardized structural Hamming distances over the $50$ replicates for Experiment 2.}
    \label{fig:clusteranalysis40}
\end{figure}

\paragraph{Experiment 2: clustered graph}
The second experiment considers a block structured  graph with different sparsity levels, namely $0.1$, $0.2$ and $0.3$, as shown in \Cref{fig:cluster_graphs}. 
We fit model \eqref{fgm_model} assuming the uniform distribution in \eqref{eqn:prior_unif} over the graph space. 
The results are summarized in \Cref{fig:clusteranalysis40}. 
Overall, the performance of recovering the true graph improves as the sparsity level increases, as expected. The posterior estimate of the graph based on the BFDR outperforms the posterior median graph, i.e., the one obtained with the $0.5$ threshold.  

\subsection{Comparison with an existing method} 
\label{sec:sim_comparison}
Here, we offer a comparison with the results obtained using the Bayesian model proposed by \citet{yang2017} for simultaneously smoothing multiple functional curves, referred to as the Bayesian framework with approximations by basis functions (BABF).
As mentioned in the Introduction,  \citet{yang2017} assumed that all functional observations come from Gaussian processes that are approximated through basis functions. 
Thus, the induced Bayesian hierarchical model for the basis expansion coefficients has the same likelihood of model  \eqref{fgm_model}. Moreover, the authors adopted a Gaussian-Wishart prior for basis expansion coefficients. In particular, their covariance matrix $\bm{\Sigma}=\bm\Omega^{-1}$ is assumed to follow, a priori, an Inverse-Wishart prior centered on a Matérn covariance function, 
i.e.,  $\bm\Sigma \sim \operatorname{IW}(d,\sigma^{2}\bm{A})$, where $\bm{A}$ is a Matérn correlation matrix. We refer the reader to \citet{lindgren2011} for a detailed discussion on the link between Gaussian processes in the Matérn class and the corresponding Gaussian Markov random fields with a sparse precision matrix.

Following \cite{yang2017}, we generate 
$200$ functional curves from a Gaussian process with mean $\mu(t) = 3\,\sin(4t)$ and covariance $\bm\Sigma(s,t) = 5\,  \mbox{Matérn}(\lvert s - t \lvert, \rho=0.5, \nu=0.5)$ 
where $s,t\in[0,\pi/2]$ and  $\mbox{Matérn}(d; \rho, \nu) = (\Gamma(\nu) 2^{\nu-1})^{-1}(\sqrt{2\nu}d/\rho)^\nu K_\nu (\sqrt{2\nu}d/\rho)$, with $\Gamma(\cdot)$  the gamma function and $K_\nu(\cdot)$ the modified Bessel function of the second kind. We fit both model  \eqref{fgm_model} and BABF to the simulated curves using $p=20$ basis functions. Easy implementation of BABF model is available in the \texttt{BDFA Matlab} toolbox \citep{BFDA}. To provide a fair comparison we match prior hyperparameters by setting $d=5$ and $\bm{D} = 5\bm{I}_{p}$ in \eqref{eqn:Gwish}. Again, we place  a uniform prior over the graph space.
We repeated the analysis on $50$ simulated datasets. In the following, we report the average indices 
as well as their standard deviations (sd).

\begin{figure}[ht]
    \centering
    \includegraphics[width=0.32\textwidth]{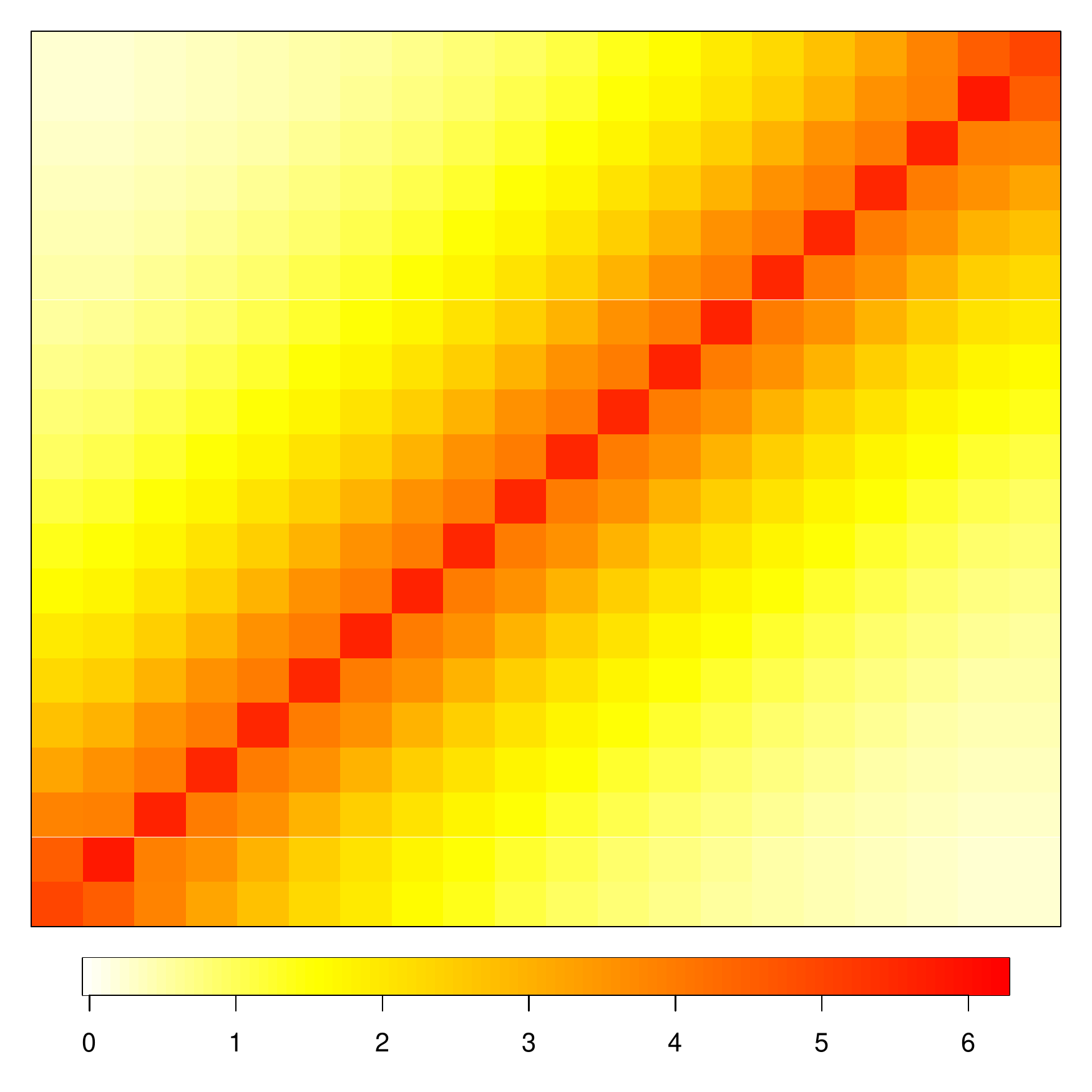}
    \includegraphics[width=0.32\textwidth]{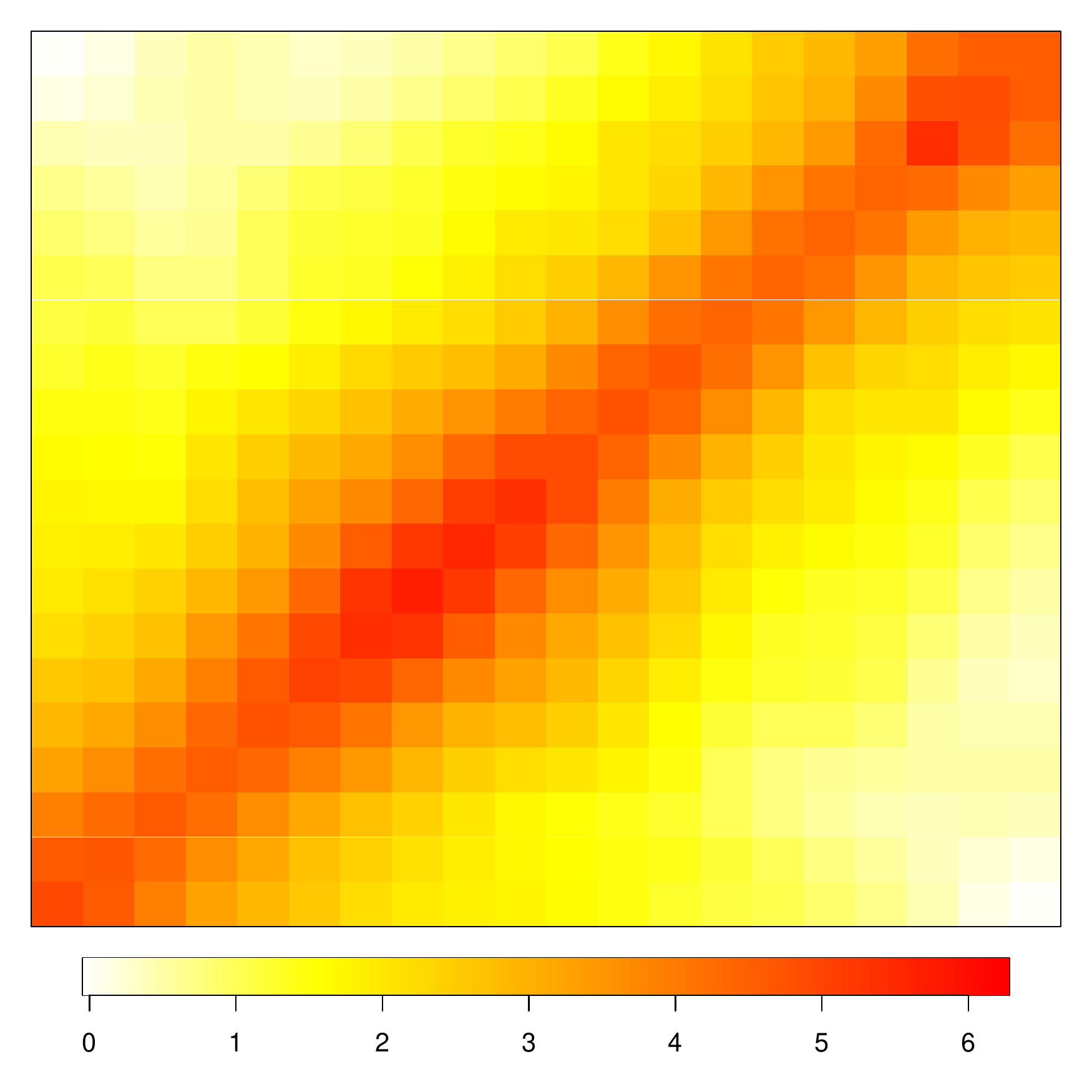}   
    \includegraphics[width=0.32\textwidth]{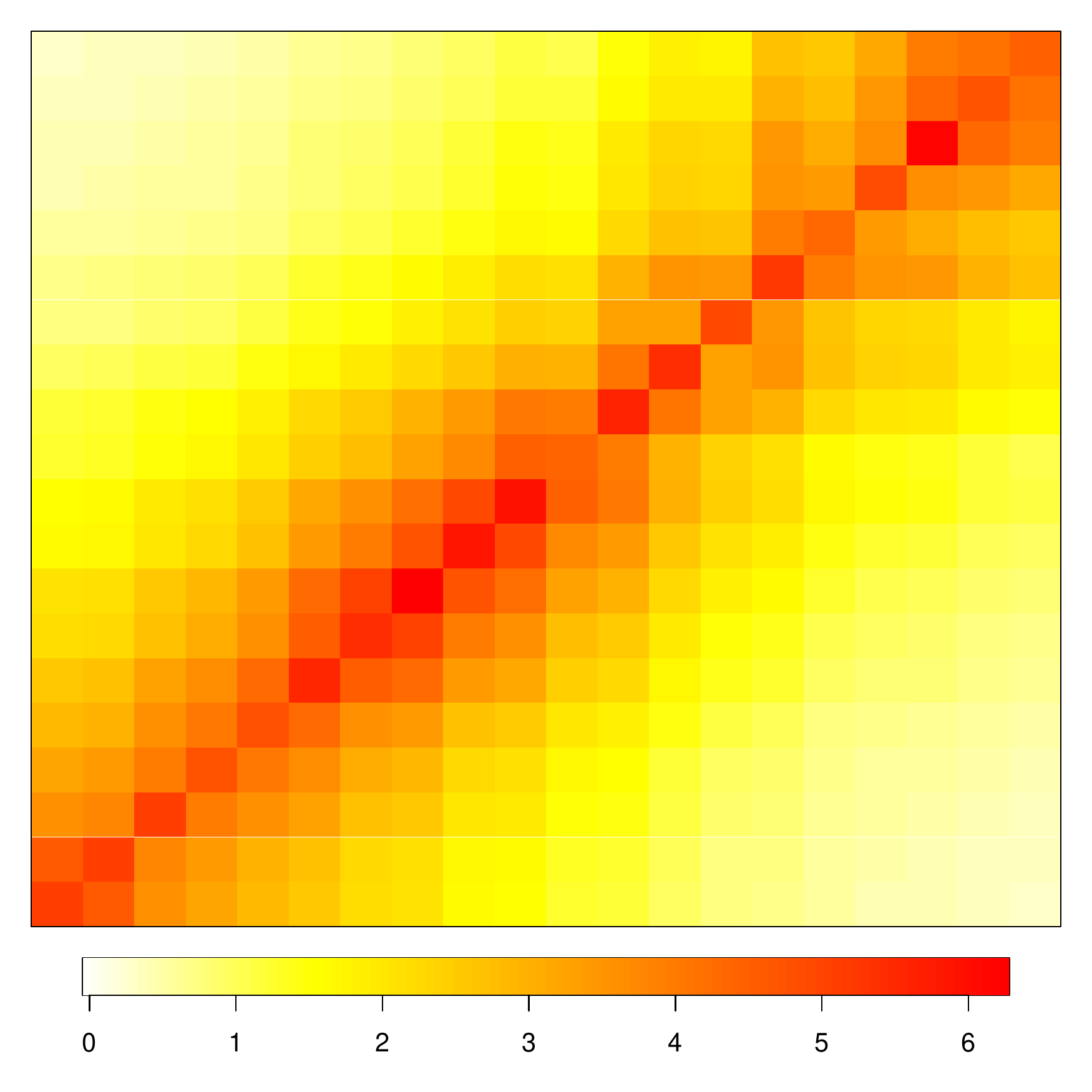}
    \caption{True covariance matrix $\bm\Sigma_{\text{true}}$ (left panel) and posterior mean of $\bm\Sigma$ obtained from BABF (middle panel) and from  model \eqref{fgm_model} (right panel). }
    \label{fig:cov}
\end{figure}

As far as concerns the smoothing performances, both methods are effective and comparable. Indeed,  
the root mean square error index is, on average, equal to $1.096$ (sd $0.019$) and $1.187$ (sd $0.013$) for our approach and BABF, respectively. 
\Cref{fig:cov} shows the true covariance matrix $\bm\Sigma_{\text{true}}$ (left panel) and the corresponding estimate $\hat{\bm\Sigma}$ obtained from BABF (middle panel) and from model \eqref{fgm_model} (right panel). 
As in \citet{mohammadi.wit2015} and \citet{wang2010}, we use the Kulbach-Leibler divergence \citep{kullback.leibler1951}
as an index to asses the distance between $\Hat{\bm\Sigma}$ and $\bm\Sigma_{\text{true}}$, that is
\begin{equation}
    \text{KL}(\bm\Sigma_{\text{true}}, \Hat{\bm\Sigma}) = \frac{1}{2}\left[\text{tr}\left(\bm\Sigma_{\text{true}}^{-1}\Hat{\bm\Sigma}\right) - p - \text{log}\left(\frac{\lvert \Hat{\bm\Sigma}\lvert}{\lvert\bm\Sigma_{\text{true}} \lvert}\right)\right].
    \label{KLdiv}
\end{equation}
The average score in equation \eqref{KLdiv} achieved by our method is $2.046$ (sd $0.9487$), which is much lower than the one obtained using BABF model, that is equal to $12.407$ (sd $1.3762$). Note that the prior distribution of $\bm\Sigma$ under BABF is centered on a Matérn structure which strongly characterizes the posterior estimate of the covariance matrix. Rather, our approach does not provide any  prior information about the underlying structure of the covariance matrix, but allows to properly learn it from the data. 

\section{Analysis of fruit purees data} \label{sec:analysis}
Food safety and authenticity is extremely important to guarantee access to good-quality and healthy products. 
According to the Food Authenticity Assurance Organisation, “food authenticity is the process of irrefutably proving that a food or food ingredient is in its original, genuine, verifiable and intended form as declared and represented”. 
Although authenticity is not a novel concept, it has recently received particular attention due to the increasing number of food frauds and adulteration. As an example, the UK National Food Crime Unit has increased from $796$ cases in 2015 to $1,193$ in 2019, with $364$ notices in the first three months of 2019\footnote{National Food Crime Unit  of the Food Standards Agency.}. 

Traditionally, all strategies and techniques developed to detect food adulterants and, more in general, to study the composition of substances, were based on chemistry. The amount of marker compounds in a test material was determined and then compared with the values obtained for previously documented authentic material. The resulting test is often time consuming and expensive. Moreover, this approach is prone towards errors, since food adulterers are becoming increasingly complex to detect. As a consequence, the pressing demand for rapid and inexpensive tests lead to the adoption of novel techniques. Nowadays, strategies based on mid-infrared spectroscopy coupled with chemometrics or statistical techniques have been used to investigate the substance composition and detect, for instance, the presence of adulterants in coffe \citep{downey1997}, fruit purees \citep{Kemsley1996,holland.kemsley.wilson1998} or minced beef \citep{MezaMrquez2010}.

\begin{figure}[ht]
    \centering
    \includegraphics{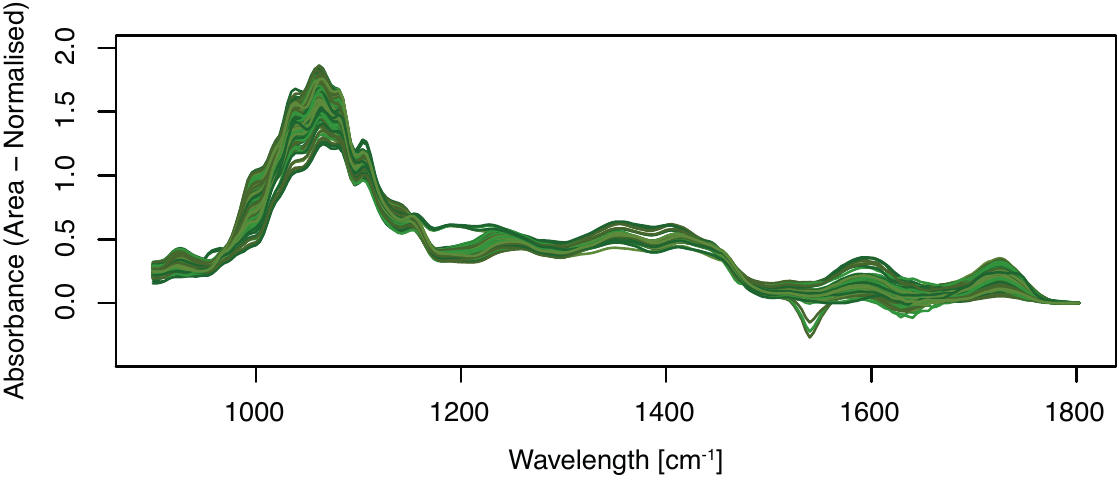}
    \caption{Plot of 351 spectra of absorbance of pure fruit purees, measured at $235$ different wavelengths of the middle-infrared spectra.}
    \label{fig:true_curves}
\end{figure}

In this work, we analyze the spectrum of absorbance of $351$ fruit purees prepared using exclusively fresh whole strawberries (without the addition of adulterants) measured on an equally-spaced grid of $235$ wavelengths and then normalized with respect to the area under the curve; see \Cref{fig:true_curves}. Data are publicly available at \url{https://data.mendeley.com/datasets/frrv2yd9rg}.
The shape of the spectra is very similar for all the curves: they are all defined by a recognizable peak at wavelengths 1000-1200 $cm^{-1}$ and few secondary others, like the ones around $1600$ $cm^{-1}$ and $1700$ $cm^{-1}$.
A further description of the data can be found in \cite{holland.kemsley.wilson1998}. 

The chemical analysis of such a complex spectrum is not trivial. Indeed we are not dealing with a pure substance that would absorb at a specific wavelength, but those purees have a more heterogeneous composition which leads to overlapping effects. The goal of the analysis is to provide useful insights about which wavelength bands are related to the different components of the purees, and more specifically about the conditional dependence structure between wavelength bands.

\begin{figure}[ht]
    \centering
    \includegraphics{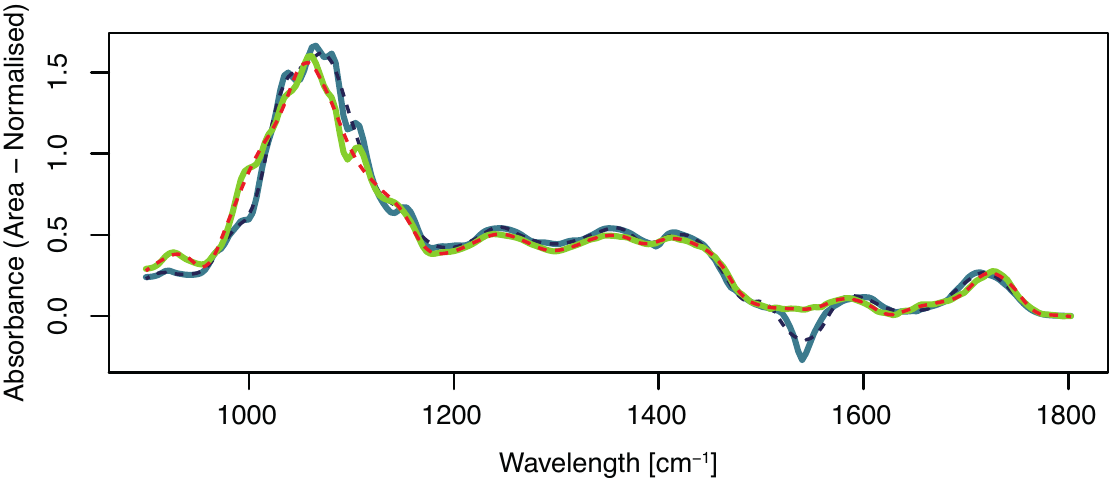}
    \caption{An example of two smoothed curves through model  \eqref{fgm_model}; solid lines represent original curves while dotted lines are the smoothed curves. }
    \label{fig:smoothed_curves}
\end{figure}

\begin{figure}[ht]
 \includegraphics[width=0.49\textwidth]{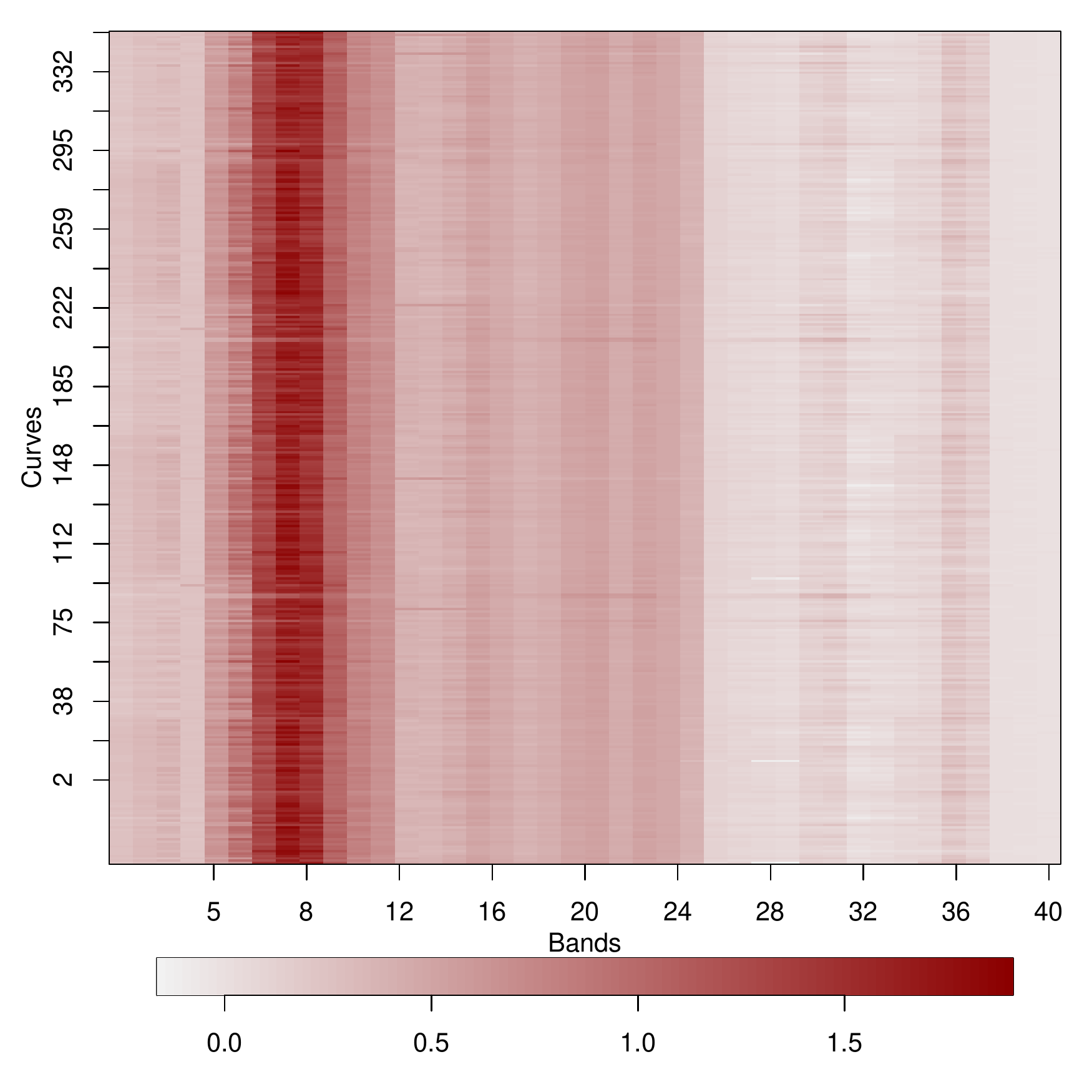}
        \includegraphics[width=0.49\textwidth]{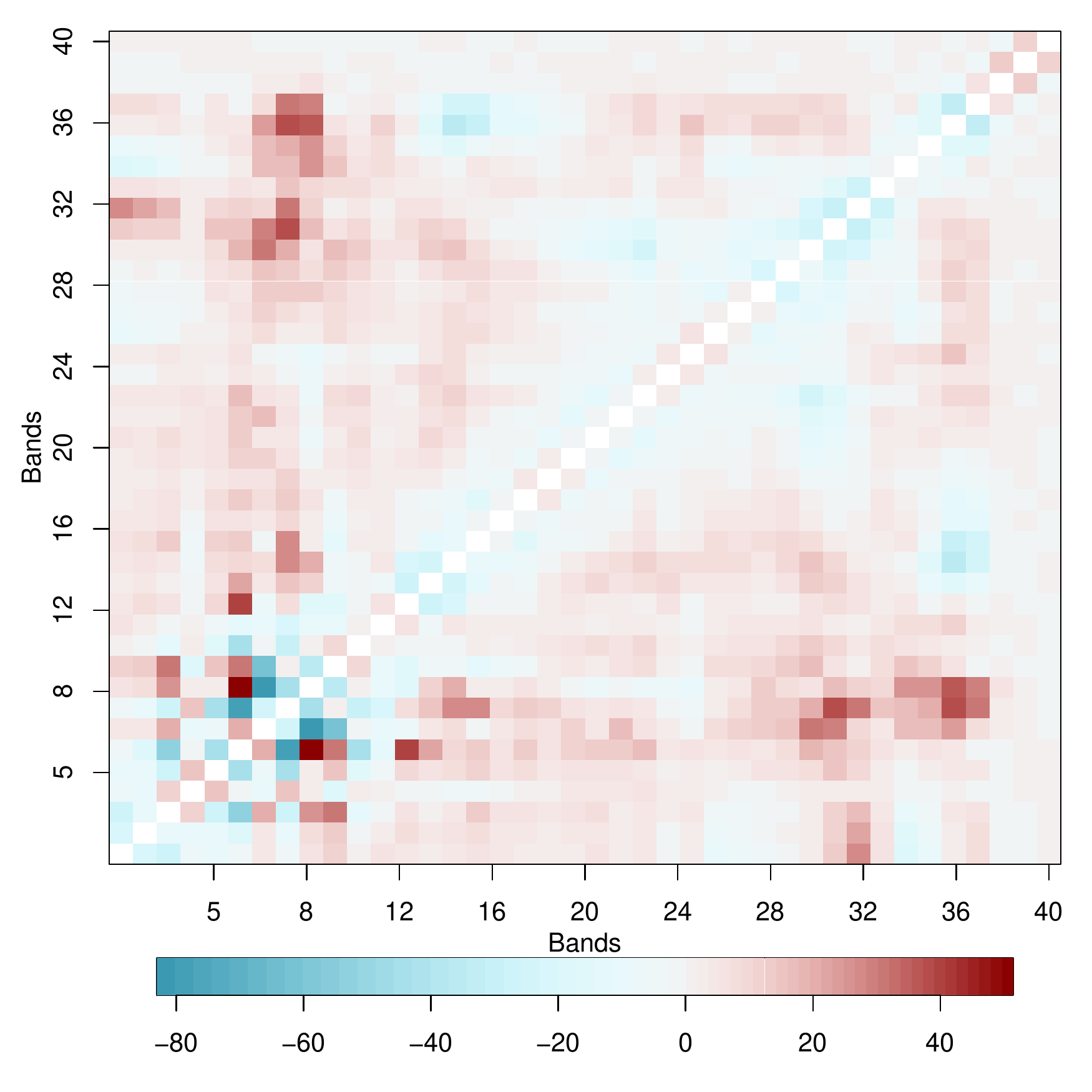}
    \caption{Posterior mean of all $\boldsymbol{\beta}$ coefficients (left panel) and the estimated precision matrix (right panel); in the latter plot, the main diagonal has been omitted.}
    \label{fig:graphical_part}
\end{figure}

We fit model \eqref{fgm_model} by assuming $p = 40$ cubic B-spline basis functions. 
Different values of $p$ were explored, obtaining similar results with respect to the ones described in this section, so they are not reported here for brevity. 
The final choice of $p=40$ faced the trade off between good fitting of the smoothed curves and limited computational burden. 
\Cref{fig:smoothed_curves} presents an example of two smoothed curves (dotted lines) compared to the original ones (solid lines). Notice that the smoothed curves follow  the shape of the original ones, smoothing away some pointwise variability. 
The left panel of \Cref{fig:graphical_part} shows the posterior mean of the coefficients while the right panel displays the estimated precision matrix $\bm{\Omega}$. 

\begin{figure}[!ht]
    \centering
    \includegraphics[width=0.49\textwidth]{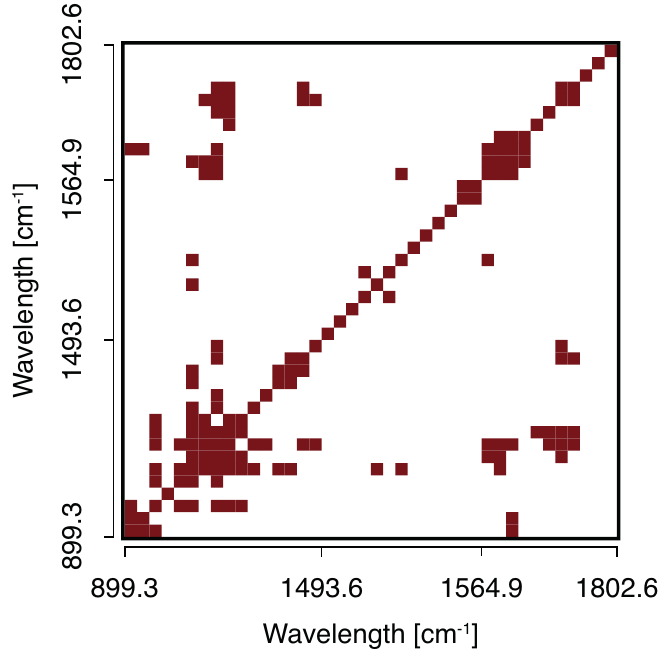}
    \hfill
    \includegraphics[width=0.49\textwidth]{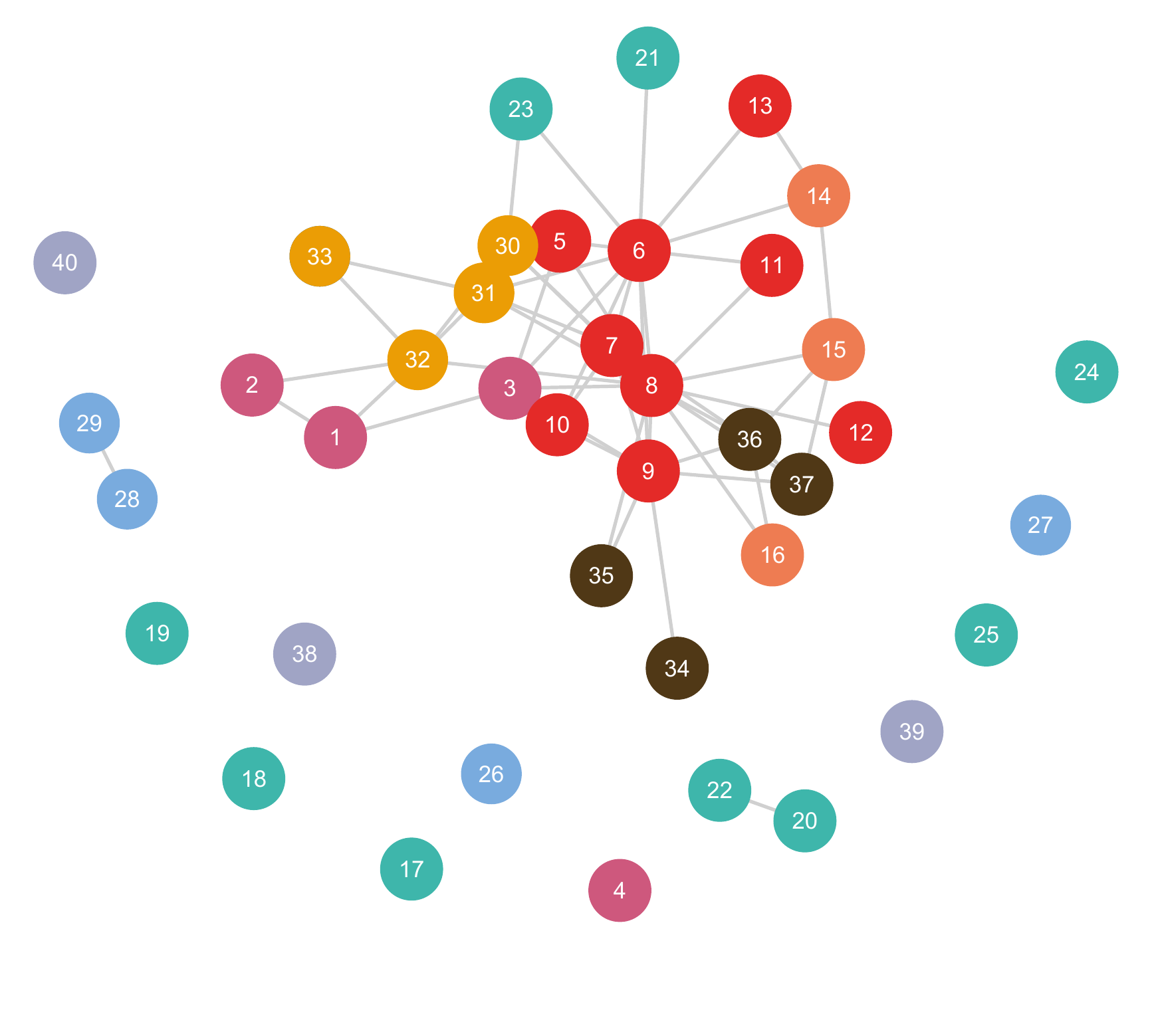}\\
    \includegraphics[width=1\textwidth]{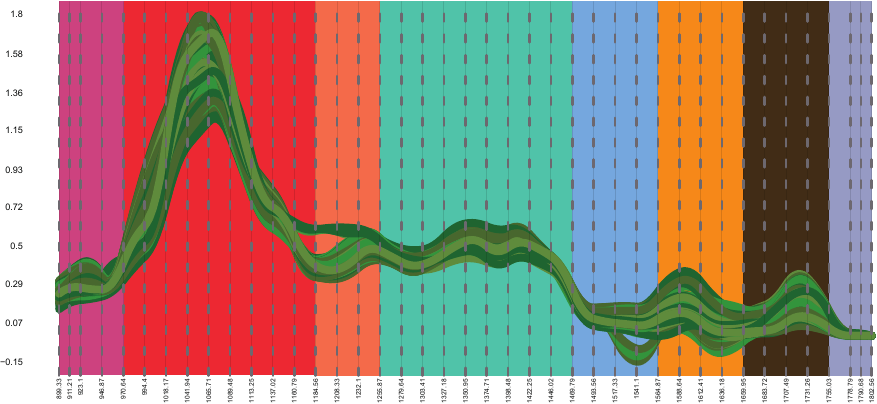}
   \caption{Top panels: posterior estimate of the graph obtained using the BFDR criterion. Nodes are colored according to different portions of the spectra highlighted in the bottom panel.}
   \label{fig:network} 
\end{figure}

The top panel of \Cref{fig:network} displays the posterior estimate of the graph obtained using the BFDR criterion.  Specifically, the top-right panel shows the estimated graph with nodes numbered and colored according to different portions of the spectra as highlighted in the bottom panel of the figure. The different colors are associated to different portions of the domain that correspond to the peaks of the signal. 
The graph offers an interpretation of the dependence structure among the wavelength bands and so the components of the purees.   
The estimated graph is quite sparse and characterized by a block structure. The  blocks are located mainly close to the diagonal, with some extra-diagonal exceptions. The prevalence of blocks close to the diagonal is expected, since close wavelength bands are likely to be associated to similar components. The blocks far away from the diagonal are  of interest, since they suggest that some components associated to very different wavelengths might be related.

The large block in the range $899.3$ - $1493.6$ $cm^{-1}$ corresponds to the main peak of the curves. The big size of this block suggests that the interactions between the corresponding bands  are not only due to the overlap of spline supports, rather they may also depend on the absorbance effects of several species overlay. 
We also notice two small off-diagonal blocks that correspond to the interactions between the bands at the highest peak and the others corresponding to lower peaks at around $1564.9$ $cm^{-1}$ and $1700$ $cm^{-1}$, respectively. 
Indeed with the infrared spectrometer, the same species can exhibit peaks at different wavelengths and our analysis identifies which peaks should be analyzed together.
Finally, looking at the graph representation, we note that  the isolated nodes in the graph are associated to intervals where the curves are rather flat (e.g., nodes 17-18-19), which means that the concentration of the corresponding substances is negligible. On the other hand, most of the identified links connects different peaks. In particular the main peak, which represents the mostly present substance, is the center of the network, all other peaks are connected to this one.


%

\section{Summary and future works} \label{sec:conclusion}
In this paper, we have 
investigated the hidden inter-dependencies among different bands of the absorbance spectrum of a given substance. To accomplish this task, we have reinterpreted the problem as a smoothing procedure of functional data by means of suitable B-spline basis functions. Our Bayesian smoothing model allows sharing of information across curves, meanwhile it 
induces  sparsity on the shared  precision matrix of the smoothing coefficients through the underlying graph structure. 

An efficient sampling strategy has been designed to provide  posterior inference as well as graphical model selection. In our simulation study, 
the selection of the graph based on  the BFDR outperformed the one based on the median summary in recovering the underlying graph. Also, we have compared the results obtained from our approach and the ones from the model proposed by \citet{yang2017}. While the two methods  are comparable in terms of smoothing, our approach outperforms the competitor in recovering the  dependence structure.


We have illustrated our methodology on a real dataset, studying the independence structure among the bands of absorbance spectrum of fruit purees. The estimated graph is sparse and equipped with a block structure; the physical interpretation of such result provides useful  insights towards the understanding of interactions of the components of the substance.




Future research will address three main challenges. 
First, since species can exhibit peaks at different wavelengths,  a clustered graph might be expected a priori. Therefore,  future works will find us in investigating a prior distribution on the graph space that induces a block structure. On the computational side, 
local moves represent a limitation of the sampling strategy, indeed the birth and death process  adds or removes only one edge at each step. 
Hence, it would be interesting to explore how to introduce also global moves for the Markov chain.
Finally, the model can be  generalized to functional discriminant  analysis in order to compare different substances and to detect adulterated ones. 


\section*{Acknowledgements}
The authors thank Guido Consonni and Alessandra Guglielmi 
for their valuable discussions and suggestions. The research of the fourth, fifth and sixth author has been partially supported by a grant from Universit\`a Cattolica del Sacro Cuore, Italy (track D1).

\bibliographystyle{elsarticle-num-names}
\bibliography{CSDA_biblio}

\appendix

\end{document}